 \documentclass [12pt,a4paper      ]{article}
\usepackage{times}

\DeclareFontFamily{OT1}{times}{}
\DeclareFontShape {OT1}{times}{m }{n }{ <-> ptmr }{}
\DeclareFontShape {OT1}{times}{bx}{n }{ <-> ptmb }{}
\DeclareFontShape {OT1}{times}{m }{it}{ <-> ptmri}{}
\DeclareFontShape {OT1}{times}{bx}{it}{ <-> ptmbi}{}
\usepackage{amsmath}
\usepackage{amsfonts}
\usepackage{amssymb}
\usepackage{latexsym}
\newcommand{\MUNU}{ {\mu\nu} }
\newcommand{\I}{ {\Upsilon} }
\newcommand{\II}{ {\DUP} }
\newcommand{\SA}{ \mathcal{A} }

\usepackage{isridef}  
\numberwithin{equation}{section}
\renewcommand{\today}{31 December 2008}

\begin{document}

\title{\bf\vspace{-2.5cm} The self-interaction force on an
                          arbitrarily moving point-charge
                          and its energy-momentum radiation rate:\\ 
                        {\it A mathematically rigorous derivation of the
                          Lorentz-Dirac equation of motion} }

\author{
         {\bf Andre Gsponer}\\
         {\it Independent Scientific Research Institute}\\ 
         {\it Oxford, OX4 4YS, United Kingdom}
       }

\date{ISRI-07-01.19 ~~ \today}

\maketitle

\begin{abstract}

The classical theory of radiating point-charges is revisited: the retarded potentials, fields, and currents are defined as nonlinear generalized functions and all calculations are made in a Colombeau algebra.  The total rate of energy-momentum radiated by an arbitrarily moving relativistic point-charge under the effect of its own field is shown to be rigorously equal to minus the self-interaction force due to that field.  This solves, without changing anything in Maxwell's theory, numerous long-standing problems going back to more than a century.  As an immediate application an unambiguous derivation of the Lorentz-Dirac equation of motion is given, and the origin of the problem with the Schott term is explained:  it was due to the fact that the correct self-energy of a point charge is not the Coulomb self-energy, but an integral over a delta-squared function which yields a finite contribution to the Schott term that is either absent or incorrect in the customary formulations.



~\\

\noindent 03.50.De ~ Classical electromagnetism, Maxwell equations

\end{abstract}

\section{Introduction}
\label{int:0}

Gauss's theorem in its various forms is a major tool in field theory because it enables to transform integrals over closed surfaces into integrals over the volume contained within them.   This enables, for example, to express Maxwell's equations in integral rather than differential form, and thus to establish precise relations between the total flux of fundamental field quantities through closed surfaces and the volume integrals of quantities that are differentially related to them.  

As long as all field quantities are smooth enough, and provided all surfaces are sufficiently regular and simple, there are no problems: Gauss's theorem and its numerous corollaries can be applied without restriction and with no great technical difficulties.  But this is not the case when there are singularities in the fields.

Consider for example the flux of energy-momentum radiated by a point-charge under the effect of its own field through a 3-surface enclosing its world-line, and the related 4-volume integral of the self-interaction force on this point-charge due to that field:  according to Gauss's theorem they should be equal --- provided all problems related to singularities are properly taken care of since the fields are infinite on the world-line.  Indeed, let $T_\MUNU$ by the symmetric energy-momentum tensor, and $^3\Sigma$ the 3-surface enclosing a 4-volume $^4\Omega$ surrounding the world-line between the proper-times $\tau_1$ and $\tau_2$.  The radiated energy-momentum flux is then
\begin{equation}\label{int:1}
    P_\MU(\Sigma) \DEF \iiint_{^3\Sigma} d^3\Sigma^\NU ~ T_\MUNU
                     = \int_{\tau_1}^{\tau_2} d\tau 
                       \iiint_{^3\Omega} d^3\Omega ~ \partial^\NU T_\MUNU,
\end{equation}
where Gauss's theorem was used to write the right-hand side.  Using the definition of $T_\MUNU$
\begin{equation}\label{int:2}
     T^\MUNU=  -\frac{1}{4\pi} ( F^{\mu\alpha} F^\NU_{\alpha} 
               -\frac{1}{4} g^\MUNU F^{\alpha\beta} F_{\alpha\beta} ),
\end{equation}
and Maxwell's equations
\begin{equation}\label{int:3}
  \frac{\partial F_{\mu\nu}}{\partial \lambda} +
  \frac{\partial F_{\lambda\mu}}{\partial \nu} +
  \frac{\partial F_{\nu\lambda}}{\partial \mu} = 0,
     \qquad \text{and} \qquad
    \partial^\nu F_\MUNU =  - 4\pi J_\MU,
\end{equation}
the right-hand side of Eq.\,\eqref{int:1} becomes
\begin{equation}\label{int:4}
           \int_{\tau_1}^{\tau_2} d\tau 
           \iiint_{^3\Omega} d^3\Omega ~\bigl(- F_\MUNU J^\NU \bigr)
      \DEF \int_{\tau_1}^{\tau_2} d\tau~ \dot{P}_\MU(\Omega) =  P_\MU(\Omega).
\end{equation}
Therefore, comparing with the left-hand side of Eq.\,\eqref{int:1}, Gauss's theorem implies
\begin{equation}\label{int:5}
       \dot{P}_\MU(\Sigma) = \dot{P}_\MU(\Omega) = - Q_\MU(\Omega),
\end{equation}
where
\begin{equation}\label{int:6}
       Q_\MU(\Omega) \DEF \iiint_{^3\Omega} d^3\Omega ~ F_\MUNU J^\NU.
\end{equation}

    Equation Eq.\,\eqref{int:5} states that the rate of energy-momentum flux through the 3-surface $^3\Sigma$ at the time $\tau$, calculated as the proper-time derivatives of either the 3-surface integral $P_\MU(\Sigma)$ or of the 4-volume integral $P_\MU(\Omega)$, is equal to minus the integral $Q_\MU(\Omega)$ of the self-interaction force density $F_\MUNU J^\NU$ over the 3-volume $^3\Omega$ enclosed by the 3-surface at that time.

     However, when $\dot{P}_\MU(\Sigma)$ and $\dot{P}_\MU(\Omega)$ are evaluated using for $F_\MUNU$ the customary Li\'enard-Wiechert field strengths and for $J_\MU$ the customary current density of an arbitrarily moving relativistic point-charge, what is not without difficulties and ambiguities since both $F_\MUNU$ and $J^\NU$ are unbounded on the world-line, one finds that they are not equal.  

  For example, following Lorentz \cite{LOREN1909-}, the self-force is usually calculated non-relativistically and then generalized to the relativistic case.  This gives
\begin{equation}\label{int:7}
       \dot{P}_\MU(\Omega)  =  \lim_{\xi \rightarrow 0}
                               \frac{2}{3} \frac{e^2}{\xi} \ddot{Z}_\MU 
                             - \frac{2}{3} e^2          \SA^2 \dot{Z}_\MU
                             - \frac{2}{3} e^2            \dddot{Z}_\MU,
\end{equation}
where the term containing the square of the acceleration $\SA^2=\ddot{Z}_\MU\ddot{Z}^\MU$, known as the `Larmor term,' is the invariant radiation rate, and the term in the third proper-time derivative of the position, Lorentz's frictional force, is known as the `Schott term.' 

    On the other hand, the energy-momentum flux rate can be found by either generalizing a non-relativistic calculation, or, following Dirac \cite{DIRAC1938-}, by a direct covariant 4-dimensional calculation.  This gives
\begin{equation}\label{int:8}
       \dot{P}_\MU(\Sigma) =  \lim_{\xi \rightarrow 0}
                              \frac{1}{2} \frac{e^2}{\xi} \ddot{Z}_\MU
                            - \frac{2}{3} e^2          \SA^2 \dot{Z}_\MU,
\end{equation}
in which the Schott term is missing and where the diverging quantity $e^2/2\xi$ is called the `Coulomb self-energy' or `electrostatic self-mass' of the point-charge, which differs from the `electromagnetic self-mass' $2e^2/3\xi$ which appears in \eqref{int:7} by a factor $4/3$.

   Apart from this notorious `$4/3$ problem,' the main difference between $\dot{P}_\MU(\Sigma)$ and $\dot{P}_\MU(\Omega)$ is that $\dot{P}_\MU(\Omega)$ is a Minkowski force, that is a genuine relativistic force satisfying the orthogonality condition
\begin{equation}\label{int:9}
       \dot{P}_\MU \dot{Z}^\MU = 0, 
\end{equation}
on account of the kinematical identities $\dot{Z}_\MU\dot{Z}^\MU=1$,  $\ddot{Z}_\MU\dot{Z}^\MU=0$, and $\dddot{Z}_\MU\dot{Z}^\MU=-\SA^2$, whereas  $\dot{P}_\MU(\Sigma)$ is not such a force because there is no Schott term.  In fact, this missing term is a major problem because, as noted by Dirac, the derivation of \eqref{int:8} is quite simple, and there is no way to get the Schott term in this calculation without making a supplementary assumption. 

  These problems, as well as other inconsistencies, have led to hundreds of papers and endless controversies which we are not going to review and discuss here.  Indeed, while theses difficulties are known, there is a large consensus that the finite part of the force \eqref{int:7} is the correct relativistic form of the self-force on an arbitrarily moving point-charge, i.e., the so-called `Abraham' or `von Laue' self-interaction force $\tfrac{2}{3} e^2 (\dddot{Z}_\MU +  \SA^2 \dot{Z}_\MU)$.  Similarly, there is a large consensus that the divergent terms in \eqref{int:8} and \eqref{int:7}  should be discarded or `renormalized', and that the equation motion of a point-charge in an external electromagnetic field $F_\MUNU^{\text{ext}}$ is
\begin{equation}\label{int:10}
           m\ddot{Z}_\MU  = \frac{2}{3} e^2 \bigl(   \dddot{Z}_\MU
                                                 + \SA^2 \dot{Z}_\MU \bigr)
                          + e F_\MUNU^{\text{ext}}     \dot{Z}^\NU,
\end{equation}
i.e., the Abraham-Lorentz-Dirac equation of motion.

    In this paper we show that the fundamental reason for discrepancies such as between \eqref{int:8} and \eqref{int:7} is mathematical, and due to two shortcomings of the customary calculations:
\begin{itemize}
      \item incomplete characterization of the singularities of the fields and current at the position of the point charge --- everything should be defined not just as distributions but as nonlinear generalized functions \cite{GSPON2004D,GSPON2008B};
      \item use of formalisms such as elementary calculus or Schwartz distribution theory in which the product of distributions is not defined --- this is a major defect since the energy-momentum tensor \eqref{int:2} is nonlinear in the fields so that all calculations should be made in a Colombeau algebra \cite{COLOM1984-,COLOM1985-,GSPON2006B,GSPON2008A}. 
\end{itemize}
     Once all singularities are properly defined, what basically consists of redefining the Li\'enard-Wiechert potential as a Colombeau nonlinear generalized function \cite{GSPON2006C, GSPON2006D}, the calculations of the self-energy-momentum radiation rate and of the self-interaction force are straightforward (but technically complicated) and the equality \eqref{int:5} is obtained for a strictly point-like electric charge in fully arbitrary relativistic motion without making any modification to Maxwell's theory.

   The importance of this result lies is the fact that it leads to the solution of a number of difficulties which have often been interpreted as suggesting that Maxwell's electrodynamics of point-like electrons is not fully consistent, and required to be modified at the fundamental level.  To illustrate that this is not necessary, it will be shown how the Lorentz-Dirac equation of motion derives from our rigorously calculated self-interaction force.

  Because the full calculation leading to the proof of the equality \eqref{int:5} for a relativistic point-charge is very long and complicated, the details are given in another paper \cite{GSPON2007B}.  Nevertheless the general physical concepts, the principle of the mathematical methods used, and the main steps in the calculation are explained in the present paper.

\section{Definitions of the fields and of the geometry}
\label{def:0}

Let $Z_\MU$ be the 4-position of an arbitrarily moving relativistic point-charge, and $X_\MU$ a point of observation.  At the point $X_\MU$ the 4-potential $A_\MU$, field $F_\MUNU$, and 4-current density $J_\MU$ are functions of the null interval $R_\MU$ between $X_\MU$ and $Y_\MU$, i.e.,
\begin{equation}\label{def:1}
    R_\MU \DEF X_\MU - Z_\MU,
   \qquad  \text{such that} \qquad
    R_\MU R^\MU = 0,
\end{equation}
as well as of the 4-velocity $\dot{Z_\MU}$, 4-acceleration  $\ddot{Z_\MU}$, and 4-biacceleration $\dddot{Z_\MU}$ of the charge, to which three invariants are associated: $\dot{Z}_\MU R^\MU, \ddot{Z}_\MU R^\MU$, and $\dddot{Z}_\MU R^\MU$.  The first one is called the retarded distance,
\begin{equation}\label{def:2}
          \xi \DEF   \dot{Z}_\MU R^\MU,
\end{equation}
which enables to introduce a `unit' null 4-vector $K_\MU$ defined as
\begin{equation}\label{def:3}
          K_\MU(\theta,\phi) \DEF R_\MU/\xi,
\end{equation}
and the so-called acceleration and biacceleration invariants defined as
\begin{equation}\label{def:4}
      \kappa \DEF  \ddot{Z}_\MU K^\MU,
  \qquad  \text{and} \qquad
        \chi \DEF \dddot{Z}_\MU K^\MU.
\end{equation}

   The three spatial variables $\{\xi, \theta, \phi \}$ and the proper-time $\tau$ define a causal coordinate system that is most appropriate for expressing the fields of a point-charge moving according to the world-line prescribed by $\dot{Z}_\MU(\tau)$.  As explained in Refs.~\cite{GSPON2006C} and \cite{GSPON2006D}, the retarded potential of such a point-charge must be written as a nonlinear generalized function, i.e., 
\begin{equation}\label{def:5}
    A_\MU = e \frac{\dot{Z}_\MU}{\xi}\ \UPS(\xi) ~\Bigr|_{\tau=\tau_r},
\end{equation}
where $\UPS(\xi)$, which is absent in the customary Li\'enard-Wiechert  formulation, is the generalized function defined as
\begin{equation} \label{def:6}
   \UPS(r) \DEF 
         \begin{cases}
         \text{undefined}   &   r < 0,\\
                      0     &   r = 0,\\
                     +1     &   r > 0,
         \end{cases}
   ~~~  ~~~ \text{and} \qquad
     \frac{d}{dr}\UPS(r) = \DUP(r),
\end{equation}
which explicitly specifies how to consistently differentiate at $\xi=0$.  
Instead of the customary Li\'enard-Wiechert field, the field strength is then \cite{GSPON2006C}
\begin{align}
\notag
   F_\MUNU =  e \Bigl( 
            \frac{K_\MU\ddot{Z}_\NU}{\xi}\UPS(\xi)
            &+ (1-\kappa\xi)\frac{K_\MU \dot{Z}_\NU}{\xi^2}
              \bigl( \UPS(\xi) - \xi \DUP(\xi) \bigr)\\
\label{def:7}
          - \frac{K_\NU\ddot{Z}_\MU}{\xi}\UPS(\xi)
            &- (1-\kappa\xi)\frac{K_\NU \dot{Z}_\MU}{\xi^2}
              \bigl( \UPS(\xi) - \xi \DUP(\xi) \bigr) \Bigr)
                   ~\Bigr|_{\tau=\tau_r},
\end{align}
which apart from the presence of the $\UPS$-function factors, has additional $\DUP$-like contributions.  As a consequence of these $\UPS$ and $\DUP$ factors,  the current density deriving from this field, i.e.,
\begin{equation}\label{def:8}
    J_\MU=  \frac{e}{4\pi}
             \Bigl(   \frac{\dot{Z}_\MU}{\xi^2} 
                    + \frac{\ddot{Z}_\MU + 2\kappa K_\MU}{\xi}
                    - (2\kappa^2 + \chi) K_\MU
              \Bigr)  \DUP(\xi) ~\Bigr|_{\tau=\tau_r} ,
\end{equation}
is significantly more complicated than the customary current density, but turns out to be locally conserved, whereas the customary one is not \cite{GSPON2006C,GSPON2006D}.  In this equation, as in Eqs.\,\eqref{def:6} and \eqref{def:7},  the condition  $\tau=\tau_r$ implies that all quantities are evaluated at the retarded proper time $\tau_r$.  In the following, for simplicity,  this condition will be specified explicitly only for the main equations.

   Having defined the field $F_\MUNU$ and the current density $J_\MU$ to be used to calculate the integrands of Eq.\,\eqref{int:1}, \eqref{int:4} and \eqref{int:6}, it remains to define the geometry of a 3-surface enclosing the world-line in order to calculates the integrals themselves.  

In view of this we begin by ignoring the fact that $F_\MUNU$ is singular on the world-line.  We are then free to chose any closed 3-surface intersecting the world-line at two fixed proper times $\tau_1$ and $\tau_2$.  This is because by Gauss's theorem the field-integrals over the 4-volume enclosed between any two such 3-surfaces will give zero since  $\partial^\NU F_\MUNU =0$, and thus $\partial^\NU T_\MUNU =0$, in that 4-volume.  To be consistent with the already defined causal coordinate system $\{\tau, \xi, \theta, \phi \}$ we therefore take a 3-surface consisting of one 3-tube of constant retarded distance $\xi = \Cst$ between the times $\tau_1$ and $\tau_2$, and, to close that 3-tube at these times, two 3-spheres $\tau = \Cst$ at the times $\tau_1$ and $\tau_2$ between the world-line and the radius of the 3-tube.  We therefore do not take a light-cone (defined as $\tau_r \DEF \tau - \xi = \Cst$) to close the tube at both ends as is often done to simplify calculations when only surface integrals are calculated: such a choice would make the corresponding 4-volume integrals significantly more complicated.  Besides, as the formulation of our fields involves $\UPS$ and $\DUP$ functions whose argument is $\xi$, taking surfaces and volumes different from those defined by constant values of the causal coordinates $\xi$ and $\tau$ would make the evaluation of the fields singularities at $\xi=0$ even more complicated than they already turn out to be.  Finally, with our choice, the 4-volume element is the simplest possible, namely
\begin{equation}\label{def:9}
       d^4\Omega = d\tau~ d\phi~ d\theta \sin(\theta)~ d\xi ~\xi^2 ,
\end{equation}
so that the 4-volume integrals are simply proper-time integrals times ordinary 3-volume integrals with the radial variable equal to $\xi$.

Now, to deal with the singularities on the world-line, we replace the just defined 4-volume, that is such that $\xi$ varies between $\xi=0$ and some maximum $\xi = \xi_2$, by a `hollow' 4-volume in which  $\xi$ varies between $\xi=\xi_1 \neq 0$ and the maximum $\xi = \xi_2$.  This 4-volume will then be comprised between two 3-tubes of radii $\xi_1$ and $\xi_2$, where $\xi_1$ will eventually tend to $0$ at the end of the calculation. Thus, we will calculate the 4-volume integral
\begin{equation} \label{def:10}
      \int_{\tau_1}^{\tau_2} d\tau \iiint d^3\Omega~ (...)
    = \int_{\tau_1}^{\tau_2} d\tau \int_{\xi_1}^{\xi_2} d\xi~ \xi^2
        \iint d\omega~ (...),
\end{equation}
in which (for practical reasons) the integration over the ordinary solid angle $d\omega = d\phi~ d\theta \sin(\theta)$ will be done first in general.

In this geometry, the normal to the 3-tube $\xi = \Cst$ is
\begin{equation}\label{def:11}
       N^T_\MU = (1-\kappa\xi) K_\MU + \dot{Z}_\MU  ~\Bigr|_{\tau=\tau_r},
\end{equation}
and the integrals over the internal and external 3-tubes are
\begin{equation}\label{def:12}
       \iiint d^3\Sigma^T_\MU~ (...) = \int_{\tau_1}^{\tau_2} d\tau
                    \iint d\omega~ \xi^2 N^T_\MU~ (...) \Bigr|_{\xi_1}^{\xi_2},
\end{equation}
where $\xi_1$ is the internal radius of the tube that is sent to zero at the end of the calculation.

Similarly, the normal to the 3-sphere $\tau = \Cst$ is
\begin{equation}\label{def:13}
       N^S_\MU =  -\kappa\xi K_\MU + \dot{Z}_\MU  ~\Bigr|_{\tau=\tau_r},
\end{equation}
and the combined integrals over the two hollow 3-spheres enclosing the hollow 4-volume at both ends can be written
\begin{equation}\label{def:14}
       \iiint d^3\Sigma^S_\MU~ (...) = \int_{\xi_1}^{\xi_2} d\xi
                   \iint d\omega~ \xi^2N^S_\MU~ (...) \Bigr|_{\tau_1}^{\tau_2}. 
\end{equation}

\section{Mathematical methodology}
\label{mat:0}

The proof of Eq.\,\eqref{int:5} involves calculating products of the generalized functions $\UPS, \DUP$, and $\DUP'$ such as $\UPS^2$, $\DUP\UPS$, $\DUP'\UPS$, $\DUP'\DUP$, and $\DUP^2$, possibly multiplied by $\xi^n$ with $n \in \mathbb{Z}$.  If these generalized functions were defined as Schwartz distributions these products would not be defined and all sorts of inconsistencies would arise.  But as we define them as nonlinear generalized functions we can use Colombeau's theory which enables to multiply, differentiate, and integrate them as if we were working with ordinary smooth functions \cite{COLOM1984-,COLOM1985-,GSPON2006B,GSPON2008A}.

   In Colombeau's theory the generalized functions $\UPS, \DUP$, etc., are embedded in an algebra by convolution with a `mollifier function' $\eta(x)$, i.e., an unspecified real smooth function whose minimal properties are given by 
\begin{equation} \label{mat:1}
     \int dz~\eta(z) = 1,
    \quad \text{and} \quad
    \int dz~z^n\eta(z) = 0,
    \quad \forall n=1,...,q \in \mathbb{N}.
\end{equation}
This function can be seen as describing details of the singularities at the infinitesimal level.  These details are possibly unknown, but not essential for getting meaningful results which are generally obtained by integrations in which only global properties such as \eqref{mat:1} are important.  However, as will be the case in this paper, further constraints can be put on the mollifier, for instance to meet requirements imposed by the physical problem investigated.

   Of course, working in a Colombeau algebra requires some precautions:  all `infinitesimal' terms aring as a result of a differentiation or any other operation must be kept until the end of the calculation --- contrary to calculating with distributions where such terms may possibly be discarded.  Similarly, distributional identities like $\DUP'(x)=-\DUP(x)/x$, which was used to write the conserved Li\'enard-Wiechert current density in the form \eqref{def:8}, have to be avoided to remain as general as possible.  We will therefore use $J_\MU$ in the form \cite[Eq.\,(4.17)]{GSPON2006D}
%
\begin{equation} \label{mat:2}
  J_\MU = \frac{e}{4\pi}\Bigl(
        - (1-2\kappa\xi)\dot{Z}_\MU          \frac{1}{\xi  }\DUP'
        + (\ddot{Z}_\MU -\xi\chi K_\MU) \frac{1}{\xi}\DUP
        + \kappa N^T_\MU          \bigl(\frac{1}{\xi}\DUP - \DUP'\bigr)
                        \Bigr),
\end{equation}
where $N^T_\MU$ is the normal to the 3-tube $\xi = \Cst$, i.e.,  Eq.\,\eqref{def:11}.

   To calculate the integrals \eqref{def:10}, \eqref{def:12}, and  \eqref{def:14}, we introduce a notation that will be systematically used in the following.  That is, we rewrite the field strength as a sum of two terms
\begin{equation}\label{mat:3}
        F_\MUNU(\tau,\xi) =   F_\MUNU^\I (\tau,\xi) \UPS(\xi)
                            + F_\MUNU^\II(\tau,\xi) \DUP(\xi), 
\end{equation}
where the upper indices `$^\I$' and `$^\II$' refer to the $\UPS$ part (i.e., the customary Li\'enard-Wiechert field) and respectively to the $\DUP$ part of Eq.\,\eqref{def:7}.  The energy-momentum tensor $T_\MUNU$ consists then of four terms, and consequently the surface integrals \eqref{def:12} and \eqref{def:14} comprise also four terms, i.e., $P_\MU^{\I,\I}(\Sigma)$, $P_\MU^{\I,\II}(\Sigma)$, $P_\MU^{\II,\I}(\Sigma)$, and $P_\MU^{\II,\II}(\Sigma)$, in which the factors $\UPS^2, \UPS\DUP, \DUP\UPS$, and $\DUP^2$ appear.  Similarly, the volume integral \eqref{def:10} involves then two terms $P_\MU^{\I,\II}(\Omega)$ and $P_\MU^{\II,\II}(\Omega)$, in which the factors $\DUP\UPS$ and $\DUP'\UPS$, and respectively $\DUP\DUP'$ and $\DUP^2$, appear.

   As for the integrations themselves, we perform them according to a similar pattern, in which all $\xi$ integrations are left for the end because both $\DUP$ and $\UPS$ are functions of $\xi$ which are singular as  $\xi \rightarrow 0$.  Thus we begin by making the angular integrations, which together with the complicated algebra implied by the products of the many field components is the most laborious part of the calculation.  This work is however greatly facilitated by the biquaternion techniques due to Paul Weiss explained in Refs.~\cite{GSPON2006D} and \cite{GSPON2007B}, which take maximum advantage of the cancellations between null 4-vectors, and which allow a maximal separation of variables and factorization of all quantities in terms of factors depending solely of kinematical or angular variables \cite{WEISS1941-}.  The resulting surface and volume integrals are given in Secs.\,\ref{sur:0} and \ref{vol:0}.

    Next we transform the resulting surface integrals in such a way that they have a form directly comparable to the volume integrals, namely a double integration over $\xi$ and $\tau$. This enables, after dropping the $\tau$ integration, to compare the remaining $\xi$-integrals, which should be equal according to Eq.\,\eqref{int:5}.  The integrands of Eqs.\,\eqref{def:12} and \eqref{def:14}, which are initially integrated over $\tau$ and respectively $\xi$ alone, have therefore to be transformed and combined in non-trivial ways.  This is made possible  by the fact that all kinematical quantities are implicit functions of the retarded time $\tau_r = \tau - \xi$, so that their $\tau$ and $\xi$ derivatives are related as
\begin{equation}\label{mat:4}
           \frac{\partial}{\partial \tau} E(\tau_r)
       = - \frac{\partial}{\partial \xi } E(\tau_r),
       \qquad \text{or} \qquad
       \dot{E}(\tau_r) = -E(\tau_r)',
\end{equation}
where the dot and the prime denote differentiation with respect to $\tau$ and to $\xi$, respectively, and where $E(\tau_r)$ is any expression function of the variable $\tau_r$.  A particularly interesting intermediate result in the transformation of the surface integrals is given as Eqs.\,(\ref{sur:7}--\ref{sur:9}).

   Finally, the volume integrals given in Sec.\,\ref{vol:0} are transformed in such a way that there is no $\DUP'$ anymore.  This is done using the identities $(\DUP^2)' = 2 \DUP\DUP'$ and $(\xi\DUP\UPS)' =  \DUP\UPS + \xi\DUP'\UPS + \xi\DUP^2$, which enable to integrate \eqref{vol:2} and \eqref{vol:3} by parts to eliminate $\DUP'$. 

   At this stage all expressions are comparable. It is then found that Eq.\,\eqref{int:1} holds exactly as it should according to Gauss's theorem, and that both $P_\MU(\Sigma)$ and $P_\MU(\Omega)$ can be written in the form (\ref{gem:3}--\ref{gem:6}).  Thus Eq.\,\eqref{int:5} holds as well and $\dot{P}_\MU(\Sigma)=\dot{P}_\MU(\Omega)$ can be written in the form (\ref{gsi:1}--\ref{gsi:4}).  However, since we have not yet made the final $\xi$ integrations, these equalities are only a check that all algebraic calculations and angular integrations are correct, and that the integrands have been properly transformed using Eq.\,\eqref{mat:4} and the rules of partial integration.  The precautions taken to properly deal with the singularities, and the postulate that everything is calculated in a Colombeau algebra, will become essential when making these $\xi$ integrations and taking the limits $\xi_2 \rightarrow \infty$ and  $\xi_1 \rightarrow 0$, what will be done in Sec.\,\ref{gsi:0}.

   Finally, until Sec.\,\ref{gsi:0}, we will take $e=1$ for the electric charge to avoid the factor $e^2$ which is multiplying all surface and volume integrals.

\section{Surface integral: the energy-momentum flow}
\label{sur:0}

The four contributions $P^{\I,\I}_\MU$,  $P^{\I,\II}_\MU$, $P^{\II,\I}_\MU$, and $P^{\II,\II}_\MU$ to the surface integral $P_\MU(\Sigma)$, i.e., Eq.\,\eqref{int:1} or the proper-time integral of the left-hand side of \eqref{int:5}, are:
%
%
\begin{align}
\label{sur:1}
  P^{\I,\I}_\MU(\Sigma) &= \int_{\tau_1}^{\tau_2} d\tau 
     ~\Bigr({\frac{1}{2\xi} \ddot{Z}_\MU
           - \frac{2}{3} \SA^2 \dot{Z}_\MU
     }\Bigl) \UPS^2 \Bigr|_{\xi_1}^{\xi_2}\\
\label{sur:2}
                      &+ \int_{\xi_1}^{\xi_2} d\xi
     ~\Bigr({  \frac{1}{2\xi^2} \dot{Z}_\MU
             + \frac{1}{2\xi  }\ddot{Z}_\MU
             - \frac{2}{3}  \SA^2 \dot{Z}_\MU
     }\Bigl) \UPS^2 \Bigr|_{\tau_1}^{\tau_2} ,
\end{align}
%
%
\begin{align}
\label{sur:3}
 P^{\I,\II}_\MU(\Sigma) + P^{\II,\I}_\MU(\Sigma) &= \int_{\tau_1}^{\tau_2} d\tau 
     ~\Bigr({ -\frac{1}{6} \SA^2 \dot{Z}_\MU
     }\Bigl) 2 \xi \DUP\UPS \Bigr|_{\xi_1}^{\xi_2}\\
\label{sur:4}
                      &+ \int_{\xi_1}^{\xi_2} d\xi
     ~\Bigr({ -\frac{1}{2\xi^2}  \dot{Z}_\MU
             - \frac{1}{6\xi}   \ddot{Z}_\MU
             - \frac{1}{6}   \SA^2 \dot{Z}_\MU
     }\Bigl) 2  \xi \DUP\UPS \Bigr|_{\tau_1}^{\tau_2} ,
\end{align}
%
%
\begin{align}
\label{sur:5}
  P^{\II,\II}_\MU(\Sigma) &= \int_{\tau_1}^{\tau_2} d\tau 
     ~\Bigr({ -\frac{  1}{2\xi}   \ddot{Z}_\MU
             + \frac{  1}{3}  \SA^2  \dot{Z}_\MU
             + \frac{\xi}{10} \SA^2 \ddot{Z}_\MU
     }\Bigl)  \xi^2 \DUP^2 \Bigr|_{\xi_1}^{\xi_2}\\
\label{sur:6}
                      &+ \int_{\xi_1}^{\xi_2} d\xi
     ~\Bigr({  \frac{1}{2\xi^2}     \dot{Z}_\MU
             - \frac{1}{6\xi}      \ddot{Z}_\MU
             + \frac{1}{6}      \SA^2 \dot{Z}_\MU
             + \frac{\xi}{10}  \SA^2 \ddot{Z}_\MU
     }\Bigl)   \xi^2 \DUP^2 \Bigr|_{\tau_1}^{\tau_2} .
\end{align}
When $\xi \rightarrow 0$ the expression inside the parentheses in \eqref{sur:1} is equal to \eqref{int:8}, i.e., the usual energy-momentum flow rate through a 3-tube surrounding the world-line \cite{DIRAC1938-,WEISS1941-}.  This illustrates how much more complicated the rigorous formalism and calculations are than the customary ones.

   These integrals can be transformed according to the methods outlined in Sec.\,\ref{mat:0} and, in particular, (\ref{sur:1}--\ref{sur:4}) can be written
%
\begin{equation}
\label{sur:7}
P^{\I,\I}_\MU(\Sigma) = \int_{\tau_1}^{\tau_2} d\tau \int_{\xi_1}^{\xi_2} d\xi
     ~\Bigr({ +\frac{1}{2\xi}  \ddot{Z}_\MU
              -\frac{2}{3}  \SA^2 \dot{Z}_\MU
     }\Bigl) 2 \DUP\Upsilon ,
\end{equation}
and
%
%
\begin{align}
\label{sur:8}
     P^{\I,\II}_\MU(\Sigma) + P^{\II,\I}_\MU(\Sigma)
                      &= \int_{\tau_1}^{\tau_2} d\tau \int_{\xi_1}^{\xi_2} d\xi
     ~\Bigr({ -\frac{1}{2\xi} \ddot{Z}_\MU
              -\frac{1}{6}   \dddot{Z}_\MU 
     }\Bigl) 2 \DUP\Upsilon \\
\label{sur:9}
                      &+ \int_{\tau_1}^{\tau_2} d\tau \int_{\xi_1}^{\xi_2} d\xi
     ~\Bigr({  -\frac{1}{6}  \SA^2 \dot{Z}_\MU
     }\Bigl) 2 (\xi \DUP\Upsilon)' .
\end{align}
When adding \eqref{sur:7} and \eqref{sur:8} a remarkable cancellation takes place: the $1/2\xi$ diverging terms cancel each other.  This means that the Coulomb self-energy does not contribute to the total surface integral $P_\MU(\Sigma)$.  Indeed, the diverging term which will be interpreted as the `self-mass' will not be the usual Coulomb self-energy, i.e., the first term in \eqref{sur:1}, but the leading divergence in (\ref{sur:6}), which is an integral over a $\DUP^2$-function.  This is consistent with what was found in Ref.\,\cite{GSPON2008B} for a point-charge at rest, namely that its self-energy is not the usual Coulomb self-energy but an integral over a $\DUP^2$-function, so that the self-mass is fully concentrated at the position of the singularity.

\section{Volume integral: minus the $\tau$-integrated self-force}
\label{vol:0}

The two contributions $P^{\I,\II}_\MU$, and $P^{\II,\II}_\MU$ to the volume integral \eqref{int:4}, i.e., minus the proper-time integral of the right-hand side of \eqref{int:5}, are:
%
%
\begin{align}
\label{vol:1}
P^{\I,\II}_\MU(\Omega) &= \int_{\tau_1}^{\tau_2} d\tau \int_{\xi_1}^{\xi_2} d\xi
     ~\Bigr({- \frac{1}{3}    \dddot{Z}_\MU
             - \frac{5}{3}  \SA^2 \dot{Z}_\MU 
     }\Bigl)  \DUP\UPS \\
\label{vol:2}
                      &+ \int_{\tau_1}^{\tau_2} d\tau \int_{\xi_1}^{\xi_2} d\xi
     ~\Bigr({ - \frac{1}{3} \SA^2 \dot{Z}_\MU
     }\Bigl)  \xi \DUP'\UPS ,
\end{align}
and
%
%
\begin{align}
\label{vol:3}
P^{\II,\II}_\MU(\Omega) &=\int_{\tau_1}^{\tau_2} d\tau \int_{\xi_1}^{\xi_2} d\xi
     ~\Bigr({ -\frac{1}{\xi}    \ddot{Z}_\MU
              +\frac{2}{3}   \SA^2 \dot{Z}_\MU
              +\frac{\xi}{5} \SA^2\ddot{Z}_\MU
     }\Bigl)  \xi^2\DUP\DUP' \\
                      &+ \int_{\tau_1}^{\tau_2} d\tau \int_{\xi_1}^{\xi_2} d\xi
\label{vol:4}
     ~\Bigr({ +\frac{1}{3}        \dddot{Z}_\MU
              +\frac{1}{3}     \SA^2  \dot{Z}_\MU
              +\frac{2\xi}{15} \SA^2 \ddot{Z}_\MU
              -\frac{\xi}{3}(\dddot{Z}_\NU \ddot{Z}^\NU)\dot{Z}_\MU
     }\Bigl)   \xi \DUP^2 .
\end{align}
These integrals are rather complicated, what is of course necessary to match the complexity of the  surface integrals.  In fact, this could be expected from the conclusion of Ref.\,\cite{GSPON2006C}, where is was explained that the self-force density on an arbitrarily moving point-charge is given by $F_{\mu\nu} J^\NU$ where $F_{\mu\nu}$ is the complete retarded field \eqref{def:7}, and where $J^\NU$ is the full locally-conserved current density \eqref{def:8} or \eqref{mat:2}.

\section{The general energy-momentum flow}
\label{gem:0}

By fully transforming the surface and volume integrals according to the methods sketched in Sec.\,\ref{mat:0},  the sums
\begin{align}
\label{gem:1}
      P_\MU(\Sigma) &=  P^{\I,\I}_\MU(\Sigma) + P^{\I,\II}_\MU(\Sigma)
                +  P^{\II,\I}_\MU(\Sigma) + P^{\II,\II}_\MU(\Sigma),\\
\label{gem:2}
      P_\MU(\Omega) &= P^{\I,\II}_\MU(\Omega) + P^{\II,\II}_\MU(\Omega),
\end{align}
can be shown to by equal, i.e., $P_\MU(\Sigma) = P_\MU(\Omega) \DEF P_\MU$ as required by \eqref{int:1}, and can both be put in the form:
%
\begin{align}
\label{gem:3}
                    P_\MU(\tau_1,\tau_2,\xi_1,\xi_2)
                     &= \int_{\tau_1}^{\tau_2} d\tau \int_{\xi_1}^{\xi_2} d\xi
     ~\Bigr({ -\frac{  1}{3}        \dddot{Z}_\MU
             - \frac{  4}{3}      \SA^2 \dot{Z}_\MU
             - \frac{\xi}{3}     (\SA^2 \dot{Z}_\MU)\dot{~}
     }\Bigl)  \DUP\UPS \\
\label{gem:4}
                      &+ \int_{\tau_1}^{\tau_2} d\tau
     ~\Bigr({ - \frac{1}{3}   \SA^2 \dot{Z}_\MU
     }\Bigl)  \xi \DUP\UPS \Bigr|_{\xi_1}^{\xi_2}  \\
\label{gem:5}
                      &+ \int_{\tau_1}^{\tau_2} d\tau
     ~\Bigr({ -\frac{  1}{2\xi}         \ddot{Z}_\MU
             + \frac{  1}{3}        \SA^2  \dot{Z}_\MU
             + \frac{\xi}{10}       \SA^2 \ddot{Z}_\MU
     }\Bigl) \xi^2 \DUP^2 \Bigr|_{\xi_1}^{\xi_2} \\
\label{gem:6}
                      &+ \int_{\tau_1}^{\tau_2} d\tau \int_{\xi_1}^{\xi_2} d\xi
     ~\Bigr({  \frac{    1}{2\xi}     \ddot{Z}_\MU
             - \frac{    1}{6}       \dddot{Z}_\MU
             + \frac{  \xi}{6}    (\SA^2 \dot{Z}_\MU)\dot{~}
             - \frac{\xi^2}{10}  (\SA^2 \ddot{Z}_\MU)\dot{~}
     }\Bigl)  \xi \DUP^2 ,
\end{align}
where all kinematical quantities are evaluated at the retarded proper time $\tau_r = \tau - \xi$.

   For convenience we refer to (\ref{gem:3}--\ref{gem:6}) as the `general energy-momentum flow' even though we could change its overall sign and call it the `general  $\tau$-integrated self-force.'  By `general,' we mean (i) that this flow is a function of the parameters $\tau_1,\tau_2,\xi_1$, and $\xi_2$; and (ii) that the function $\UPS$, from which $\DUP$ derives by differentiation, belongs to a class of the nonlinear generalized function which may be further specified by physical conditions leading, for instance, to additional constraints on the Colombeau mollifier $\eta$.


\section{The general self-interaction force}
\label{gsi:0}

By differentiating $P_\MU(\tau_1,\tau_2,\xi_1,\xi_2)$ with respect to the proper time, that is by suppressing the $\tau$ integrations in (\ref{gem:3}--\ref{gem:6}), we get the equality \eqref{int:5}, i.e., $\dot{P}_\MU(\Sigma) = \dot{P}_\MU(\Omega) = - Q_\MU$ with
%
\begin{align}
\label{gsi:1}
               Q_\MU(\xi_1,\xi_2) &=
  ~e^2\Bigr({   \frac{1}{3}   \SA^2 \dot{Z}_\MU
     }\Bigl)  \xi \DUP\UPS \Bigr|_{\xi_1}^{\xi_2}  \\
\label{gsi:2}
                      &+ 
  ~e^2\Bigr({  \frac{  1}{2}            \ddot{Z}_\MU
             - \frac{\xi}{3}        \SA^2  \dot{Z}_\MU
             - \frac{\xi^2}{10}     \SA^2 \ddot{Z}_\MU
     }\Bigl) \xi \DUP^2 \Bigr|_{\xi_1}^{\xi_2} \\
\label{gsi:3}
                      &+  ~e^2 \int_{\xi_1}^{\xi_2} d\xi
    ~\Bigr({  \frac{  1}{3}        \dddot{Z}_\MU
             + \frac{  4}{3}      \SA^2 \dot{Z}_\MU
             + \frac{\xi}{3}     (\SA^2 \dot{Z}_\MU)\dot{~}
     }\Bigl)  \DUP\UPS \\
\label{gsi:4}
                      &+  ~e^2 \int_{\xi_1}^{\xi_2} d\xi
  ~e^2\Bigr({- \frac{1    }{2}        \ddot{Z}_\MU
             + \frac{\xi  }{6}       \dddot{Z}_\MU
             - \frac{\xi^2}{6}    (\SA^2 \dot{Z}_\MU)\dot{~}
             + \frac{\xi^3}{10}  (\SA^2 \ddot{Z}_\MU)\dot{~}
     }\Bigl) \DUP^2 ,
\end{align}
where all kinematical quantities are evaluated at the retarded proper time $\tau_r = \tau - \xi$, and where in view of applications we have we have made the $e^2$ factor explicit.

   Equation (\ref{gsi:1}--\ref{gsi:4}) is the main result of this paper: it is an exact closed-form expression giving, in the limits $\xi_2 \rightarrow \infty$ and  $\xi_1 \rightarrow 0$, the total self-interaction force on an arbitrarily moving relativistic point-charge.  Because it leads to the Abraham-von~Laue self-interaction force when taking these limits we call it the `general self-interaction force.'

  To calculate the limits $\xi_2 \rightarrow \infty$ and  $\xi_1 \rightarrow 0$ in (\ref{gsi:1}--\ref{gsi:4}) we first remark that if $Z(\tau)$ is indefinitely differentiable in $\tau$, which implies that the velocity, the acceleration, and the biacceleration are smooth functions of $\tau$, the kinematical expressions inside the big parentheses are all smooth as functions of $\tau_r$, that is as functions of $\xi$ as well.  Then, because
\begin{equation}\label{gsi:5}
       \xi \DUP\UPS \Bigr|_{0}^{\infty} = 0,
  \qquad  \text{and} \qquad
         \xi \DUP^2 \Bigr|_{0}^{\infty} = 0,
\end{equation}
the boundary terms (\ref{gsi:1}--\ref{gsi:2}) give zero. As for  (\ref{gsi:3}--\ref{gsi:4}) we use the formulas \cite{GSPON2008B,GSPON2006B}
\begin{align}
  \label{gsi:6} 
  \int_0^\infty dr~ \UPS(r)\DUP(r) T(r) &= \frac{1}{2} T(0),\\
 \label{gsi:7}
  \int_0^\infty dr~ \DUP^2(r)T(r) 
         &= \lim_{\epsilon \rightarrow 0} C_{[0]}\frac{T(0)}{\epsilon}
                                        + C_{[1]} {T'(0)},
\end{align}
where $C_{[0]}$ and $C_{[1]}$ are functions of the Colombeau mollifier $\eta$, i.e., 
\begin{align}
\label{gsi:8} 
  C_{[0]} = \int_{-\infty}^{+\infty} dx~   \eta^2(-x),
            \qquad \text{and} \qquad
  C_{[1]} =  \int_{-\infty}^{+\infty} dx~ x \, \eta^2(-x).
\end{align}
In  (\ref{gsi:6}--\ref{gsi:7}) the function $T(r)$ is a smooth function with compact support, which implies that $T(\pm\infty)=0$, what is the case of the kinematical expressions in (\ref{gsi:3}--\ref{gsi:4}) if we suppose that $\dot{Z}(\pm\infty)=0$, i.e., that the velocity is zero at $\tau = \pm\infty$.

  Applying  (\ref{gsi:5}--\ref{gsi:7}) to  (\ref{gsi:1}--\ref{gsi:4}) we finally get for the total self-interaction force $Q_\MU^{\text{self}} \DEF Q_\MU(0,+\infty)$ the expression
\begin{equation}\label{gsi:9}
  Q_\MU^{\text{self}}(\tau) = e^2 \Bigl( 
                   - \lim_{\epsilon \rightarrow 0}
                     \frac{C_{[0]}}{2\epsilon}\ddot{Z}_\MU(\tau)
                   + \frac{C_{[1]}}{2}      {\dddot{Z}_\MU(\tau)}
                   + \frac{   1}{6}          \dddot{Z}_\MU(\tau)
                   + \frac{   2}{3}   \SA^2(\tau)\dot{Z}_\MU(\tau)
                                    \Bigr) ,
\end{equation}
where all kinematical quantities are evaluated at $\xi=0$, i.e., on the world-line at the proper-time $\tau$, and where we used \eqref{mat:4} to replace the $\xi$-differentiation in \eqref{gsi:7} by a $\tau$-differentiation to get the term $+\tfrac{1}{2} C_{[1]} {\dddot{Z}_\MU(\tau)}$.  We therefore see that the $\DUP^2$-integral \eqref{gsi:4} contributes two terms to the self-interaction-force \eqref{gsi:9}:  a diverging contribution which can be interpreted as a self-mass as in the customary calculations, and a finite additional contribution to the Schott term which does not arise in the customary calculations.  If these contributions do not need to be explicitly shown, one can rewrite \eqref{gsi:9} using \eqref{gsi:7} in the more concise form
\begin{equation}\label{gsi:10}
    Q_\MU^{\text{self}}(\tau) = e^2 \Bigl( 
    - \frac{1}{2} \int_0^\infty \DUP^2(\xi) \ddot{Z}_\MU(\tau_r)~d\xi
                  + \frac{1}{6}           \dddot{Z}_\MU(\tau)
                  + \frac{2}{3}    \SA^2(\tau)\dot{Z}_\MU(\tau)
                                     \Bigr) ,
\end{equation}
which like (\ref{gem:3}-\ref{gem:6}) is valid in any algebra of nonlinear generalized functions.

\section{The Lorentz-Dirac equation of motion}
\label{lor:0}

An immediate application of the general formalism developed in this paper and of the self-interaction force $Q^{\text{self}}_\MU$ is the derivation of the equation of motion of a point-particle in arbitrary motion in an external electromagnetic field $F_\MUNU^{\text{ext}}$.  Indeed, the total force on a point-particle under the effect of its own field \eqref{def:7} and of an external field is
\begin{align}\label{lor:1}
   m_0 \ddot{Z}_\MU(\tau)  = \iiint_{^3\Omega} d^3\Omega ~ 
                  \bigl( F_\MUNU +  F_\MUNU^{\text{ext}} \bigr)  J^\NU
                    =  Q_\MU^{\text{self}}(\tau)
                    + e F_\MUNU^{\text{ext}}(Z) \dot{Z}^\NU(\tau),
\end{align}
with the subsidiary condition
\begin{equation}\label{lor:2}
   \dot{Z}_\MU \dot{Z}^\MU = 1,
\end{equation}
which specifies that the world-line $Z(\tau)$ is the world-line of a point in Minkowski space.  In writing \eqref{lor:1} we postulated that the total force could be written $m_0 \ddot{Z}_\MU$ where $m_0$ is a constant called the `non-renormalized' mass, and that the external field $F_\MUNU^{\text{ext}}(X)$ is a smooth function of $X$ --- or at least smooth at any point $X$ near or on the world-line $Z(\tau)$ --- so that the volume integral of the external force density, i.e., $F_\MUNU^{\text{ext}}(X) J^\NU$ with $J^\NU$ given by \eqref{def:8} or \eqref{mat:2}, reduced to $eF_\MUNU^{\text{ext}}(Z) \dot{Z}^\NU$ after an elementary integration over a three-dimensional $\delta$-function.

   Substituting  $Q^{\text{self}}$ from \eqref{gsi:9} we get therefore
\begin{equation}\label{lor:3}
   \Bigl( m_0 + \lim_{\epsilon \rightarrow 0}
     \frac{C_{[0]}}{2\epsilon} \Bigr) \ddot{Z}_\MU
                   =  e^2 \Bigl(
                     \frac{C_{[1]}}{2}   {\dddot{Z}_\MU}
                   + \frac{   1}{6}       \dddot{Z}_\MU
                   + \frac{   2}{3}     \SA^2 \dot{Z}_\MU \Bigr)
                   + e F_\MUNU^{\text{ext}} \dot{Z}^\NU,
\end{equation}
where we have put the diverging self-mass term on the left-hand side.  We now use the subsidiary condition condition \eqref{lor:2}, which by differentiation leads to the identities $\ddot{Z}_\MU\dot{Z}^\MU=0$ and $\dddot{Z}_\MU\dot{Z}^\MU=-\SA^2$.  Thus, contracting \eqref{lor:3} with $\dot{Z}^\MU$ and using the antisymmetry of the the external field we obtain  
\begin{equation}\label{lor:4}
                     \frac{C_{[1]}}{2}
                   + \frac{   1}{6}
                   = \frac{   2}{3}.
\end{equation}
This equation is satisfied if the Colombeau mollifier $\eta$ is such that the moment
\begin{align}\label{lor:5} 
  C_{[1]} =  \int_{-\infty}^{+\infty} dx~ x \, \eta^2(-x) = 1,
\end{align}
which is an additional condition supplementing the minimal conditions \eqref{mat:1}. Consequently,  Eq.\,\eqref{lor:3} becomes
\begin{equation}\label{lor:6}
   \Bigl( m_0 + \lim_{\epsilon \rightarrow 0}
     \frac{C_{[0]}}{2\epsilon} \Bigr) \ddot{Z}_\MU
                 =  \frac{2}{3} e^2( \dddot{Z}_\MU + \SA^2\dot{Z}_\MU )
                   + e F_\MUNU^{\text{ext}} \dot{Z}^\NU,
\end{equation}
which after substituting the `renormalized' or `experimental' mass
\begin{equation}\label{lor:7}
   m \DEF m_0 + \lim_{\epsilon \rightarrow 0}
      \frac{C_{[0]}}{2\epsilon},
\end{equation}
is known as the Lorentz-Dirac equation of motion, i.e., Eq.\,\eqref{int:10}, where everything is evaluated at the proper time $\tau$.

   To fully appreciate what has been achieved, let us compare our derivation  to a particularly elegant approach due to Barut \cite{BARUT1974-}.  The idea is to calculate the self-interaction force by analytically continuing the Lorentz force of a point-charge on itself, i.e., $e F_\MUNU^{\text{self}} \dot{Z}^\NU(\tau)$ where $ F_\MUNU^{\text{self}}$ is the customary Li\'enard-Wiechert field, to a point $X=Z(\tau+\epsilon)$ close to the world-line.  If this done, one easily finds \cite[Eq.\,8]{BARUT1974-}
\begin{equation}\label{lor:8}
    Q^{\text{self}}_{\text{Barut}}(\tau) = e^2 \Bigl( 
                   - \lim_{\epsilon \rightarrow 0}
                     \frac{1}{2\epsilon}\ddot{Z}_\MU(\tau)
                   + \frac{   1}{6}          \dddot{Z}_\MU(\tau)
                   + \frac{   2}{3}   \SA^2(\tau)\dot{Z}_\MU(\tau) \Bigr),
\end{equation}
which essentially differs from \eqref{gsi:9} by the absence of a finite contribution equivalent to $\tfrac{1}{2}C_{[1]}\dddot{Z}_\MU(\tau)$ so that there is a problem in getting the correct Schott term.  In fact, comparing with \eqref{gsi:10}, we see that the fundamental difference is that in Barut's approach the divering mass term is the `Coulomb self-energy,' whereas in our rigorous formalism it is a `$\DUP^2$-integral.'  A similar situation arises in Dirac's approach \cite{DIRAC1938-}, in which instead of calculating the self-interaction force the flow of energy-momentum out of a 3-surface surrounding the world-line is calculated, i.e., Eq.\,\eqref{int:8}, which implies
\begin{equation}\label{lor:9}
    Q^{\text{self}}_{\text{Dirac}}(\tau) = e^2 \Bigl( 
                   - \lim_{\epsilon \rightarrow 0}
                     \frac{1}{2\epsilon}\ddot{Z}_\MU(\tau)
                   + \frac{   2}{3}   \SA^2(\tau)\dot{Z}_\MU(\tau) \Bigr),
\end{equation}
in which there is no $\dddot{Z}_\MU(\tau)$ contribution at all.

  Therefore, with self-interaction forces like \eqref{lor:8} or \eqref{lor:9}, there is no other possibility than to make one or more additional suppositions in order to get the Lorentz-Dirac equation of motion.  In a way or another this amounts to modifying or supplementing Maxwell's theory and/or the notion of causality which leads to the retarded potentials, what is not necessary with the self-interaction force \eqref{gsi:10} in which the $\DUP^2$-integral supplies the missing $\dddot{Z}_\MU(\tau)$ contribution.

\section{References}
\label{biblio}


\begin{enumerate}

\bibitem{LOREN1909-} H.A. Lorentz, The Theory of Electrons (B.G. Teubner, Leipzig, 1909) p.\,49 and pp.\,251--254.  Lorentz's calculation is reviewed in  J.D. Jackson, Classical Electrodynamics (J. Wiley \& Sons, New York, second edition, 1975) pp.\,786--790.

\bibitem{DIRAC1938-} P.A.M. Dirac, \emph{Classical theory of radiating electrons}, Proc. Roy. Soc. {\bf A 167} (1938) 148--169.


\bibitem{GSPON2004D} A. Gsponer, \emph{Distributions in spherical coordinates with applications to classical electrodynamics}, Eur. J. Phys. {\bf 28} (2007) 267--275; Corrigendum Eur. J. Phys. {\bf 28} (2007) 1241. e-print arXiv:physics/0405133.

\bibitem{GSPON2008B} A. Gsponer, \emph{The classical point-electron in Colombeau's theory of generalized functions}, J. Math. Phys. {\bf 49} (2008) 102901 \emph{(22 pages)}. e-print arXiv:0806.4682.


\bibitem{COLOM1984-} J.F. Colombeau, New Generalized Functions and Multiplication of Distributions, North-Holland Math.~Studies {\bf 84} (North-Holland, Amsterdam, 1984) 375~pp.

\bibitem{COLOM1985-} J.F. Colombeau, Elementary Introduction to New Generalized Functions, North-Holland Math.~Studies {\bf 113} (North Holland, Amsterdam, 1985) 281~pp. 


\bibitem{GSPON2006B} A. Gsponer, \emph{A concise introduction to Colombeau generalized functions and their applications to classical electrodynamics}, Eur. J. Phys. {\bf 30} (2009) 109--126. e-print arXiv:math-ph/0611069.

\bibitem{GSPON2008A} A. Gsponer, \emph{The sequence of ideas in a re-discovery of the Colombeau algebras}, Report ISRI-08-01 (2008) 28\,pp. e-print arXiv:0807.0529. 


\bibitem{GSPON2006C} A. Gsponer, \emph{The locally-conserved current of the Li\'enard-Wiechert field}, (December, 2006) 8\,pp. e-print arXiv:physics/0612090.

\bibitem{GSPON2006D} A. Gsponer, \emph{Derivation of the potential, field, and locally-conserved charge-current density of an arbitrarily moving point charge}, (2008) 20\,pp. e-print arXiv:physics/06012232.

\bibitem{GSPON2007B} A. Gsponer, \emph{Derivation of the self-interaction force on an arbitrarily moving point-charge and of its related energy-momentum radiation rate: The Lorentz-Dirac equation of motion in a Colombeau algebra} (2008) 35\,pp. e-print arXiv:0812.4812.


\bibitem{WEISS1941-} P. Weiss, \emph{On some applications of quaternions to restricted relativity and classical radiation theory}, Proc.  Roy.  Irish.  Acad. {\bf 46} (1941) 129--168.

\bibitem{BARUT1974-} A.O. Barut, \emph{Electrodynamics in terms of retarded fields}, Phys. Rev. D {\bf 10} (1974) 3335--3336.


\end{enumerate}
\end{document}